\documentclass[graybox]{svmult}

\newcommand{\be}{\begin{eqnarray}}
\newcommand{\ee}{\end{eqnarray}}
\usepackage{cite}
\usepackage{mathptmx}       
\usepackage{helvet}         
\usepackage{courier}        
\usepackage{type1cm}        
\usepackage{makeidx}         
\usepackage{graphicx}        
\usepackage{multicol}        
\usepackage[bottom]{footmisc}

\makeindex             

\begin{document}

\title*{Effects of Epistasis and Pleiotropy on Fitness Landscapes}
\author{Bj\o rn \O stman}
\institute{Bj\o rn \O stman \at Department Microbiology and Molecular Genetics \& BEACON Center for the Study of Evolution in Action, Michigan State University, \email{ostman@msu.edu}}
%
%
\maketitle

\abstract{The factors that influence genetic architecture shape the structure of the fitness landscape, and therefore play a large role in the evolutionary dynamics. Here the NK model is used to investigate how epistasis and pleiotropy -- key components of genetic architecture -- affect the structure of the fitness landscape, and how they affect the ability of evolving populations to adapt despite the difficulty of crossing valleys present in rugged landscapes. Populations are seen to make use of epistatic interactions and pleiotropy to attain higher fitness, and are not inhibited by the fact that valleys have to be crossed to reach peaks of higher fitness.}

\section{Introduction}
\label{sec:1}
Rugged fitness landscapes have been suggested to put a halt to adaptation~\cite{Gavrilets2004,Whitlock1995}. When there are multiple peaks in the fitness landscape, populations must cross valleys to achieve higher fitness, but because crossing a valley implies that organisms will have lower fitness, they are likely to get stuck on local fitness peaks. Several solutions have been proposed to the problem of valley-crossing, including non-static landscapes~\cite{Mustonen2009,Whitlock1997,Whitlock1995}, subpopulations crossing by drift~\cite{Wright1932}, and circumventing valleys by using neutral ridges~\cite{Gavrilets1997}. Here we will see that populations can indeed cross fitness valleys as long as the mutation-supply rate (product of population size and mutation rate) is not unrealistically low. When the supply of mutations is large enough, some organisms can endure lower fitness and still manage to reproduce, thereby giving them a chance to ascend adjacent fitness peaks. An initially maladapted population will most often climb the nearest peak, and if this peak happens not to be the global peak, it can then achieve higher fitness by relying on the stochastic nature of evolution. The more rugged a landscape is, the harder it becomes to cross valleys, but it turns out that not only can populations overcome relatively high levels of ruggedness, but high ruggedness also implies that the global peak is higher, leading to more efficient adaptation. Epistatic interactions between genes therefore not only constrain adaptation when the population gets stuck on a local peak, they actually boost it. Consequently, deleterious mutations are seen not as a hindrance to adaptation, but as a necessary component without which adaptation would grind to a halt.

The structure of the fitness landscape and the ruggedness that it exhibits are shaped by the interactions between the genotypes and the mutations that take place when a descending organism moves between neighboring genotypes. Epistasis and pleiotropy have distinct but related effects whereby the fitness landscape acquire a structure that either inhibits or enables an evolving population to attain higher fitness. Evolutionary dynamics is thus largely determined by three key parameters: population size, mutation rate, and the fitness landscape. With adequate information about the fitness landscape, and the factors that underlie genetic architecture, the extent to which populations can successfully utilize deleterious mutations and locate the fittest genotype can be assessed. Results from simulations of evolving populations in rugged fitness landscapes are here presented showing that adaptation is not slowed down in moderately rugged landscapes, but rather allows populations to attain higher fitness than in landscapes with no epistasis. Three hypotheses concerning epistasis and pleiotropy springing from the model employed in this work are presented for future investigation.

\section{NK Model}
To investigate the effect of epistatic interactions on the adaptive process, we employ the NK model, which is a simple system previously used to study interactions between loci with different alleles. The NK model consists of $N$ loci in circular, binary sequences~\cite{Kauffman1993,Kauffman1987}. Each of the $N$ loci contributes to the fitness of the organism via an interaction with $K$ adjacent loci. For each locus $i$ a lookup-table consisting of uniform random numbers represents the fitness component   of a binary sequence of length $K+1$. For example, $K=1$ (interaction with one other locus) is modeled by creating random numbers for the four possible binary pairs 00,01,10,11 for each of the $N$ loci, that is, the fitness component at one locus is  conditional on the allele at one other locus. The overall fitness of an organism is usually given by the average of the $N$ fitness components, but here we use the geometric mean (motivated by the fact that one could then introduce lethal mutations by setting one or more elements in the lookup-tables to zero):

\be
W=\left( \prod_i^N \omega_i \right)^{1/N}.
\ee

\begin{figure}[htp]
\begin{center}
\includegraphics[width=4.3in,angle=0]{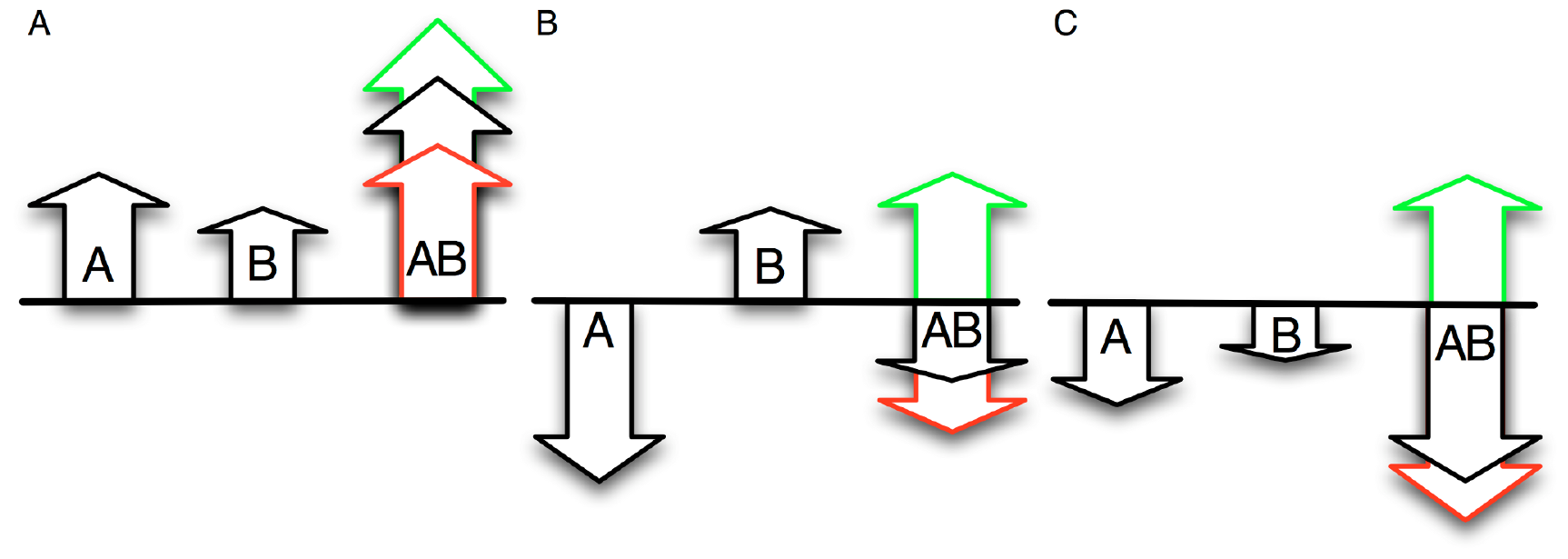}
\caption{Illustration of epistasis. (A) Two beneficial mutations occurring together in one organism have an expected non-epistatic value equal to the product of individual fitness, $W_A \times W_B = W_{AB}$ (black arrows). Fitness values higher than this correspond to positive epistasis (green arrow), and fitness values lower correspond to negative epistasis (red arrow). (B) Sign epistasis denotes situations where the effect of a mutation chances depending on the genetic background on which it occurs. Here mutation $B$ is beneficial when occurring alone, and the expectation is that the effect of $A$ and $B$ together is dominated by the larger effect size of the deleterious mutation $A$. When the actual combined effect changes from the expected deleterious to beneficial (green arrow), we observe sign epistasis. (C) Two deleterious mutations have the non-epistatic expectation of being deleterious. Reciprocal sign epistasis occurs when the combined action of the two mutations reverses the effect on fitness. Reciprocal sign epistasis occurs when a valley in the fitness landscape is crossed, and it is a necessary condition for multiple peaks to exist. From~\cite{ostman2012}.}
\label{fig_1}
\end{center}
\end{figure}

\begin{figure}[htp]
\begin{center}
\includegraphics[width=4.3in,angle=0]{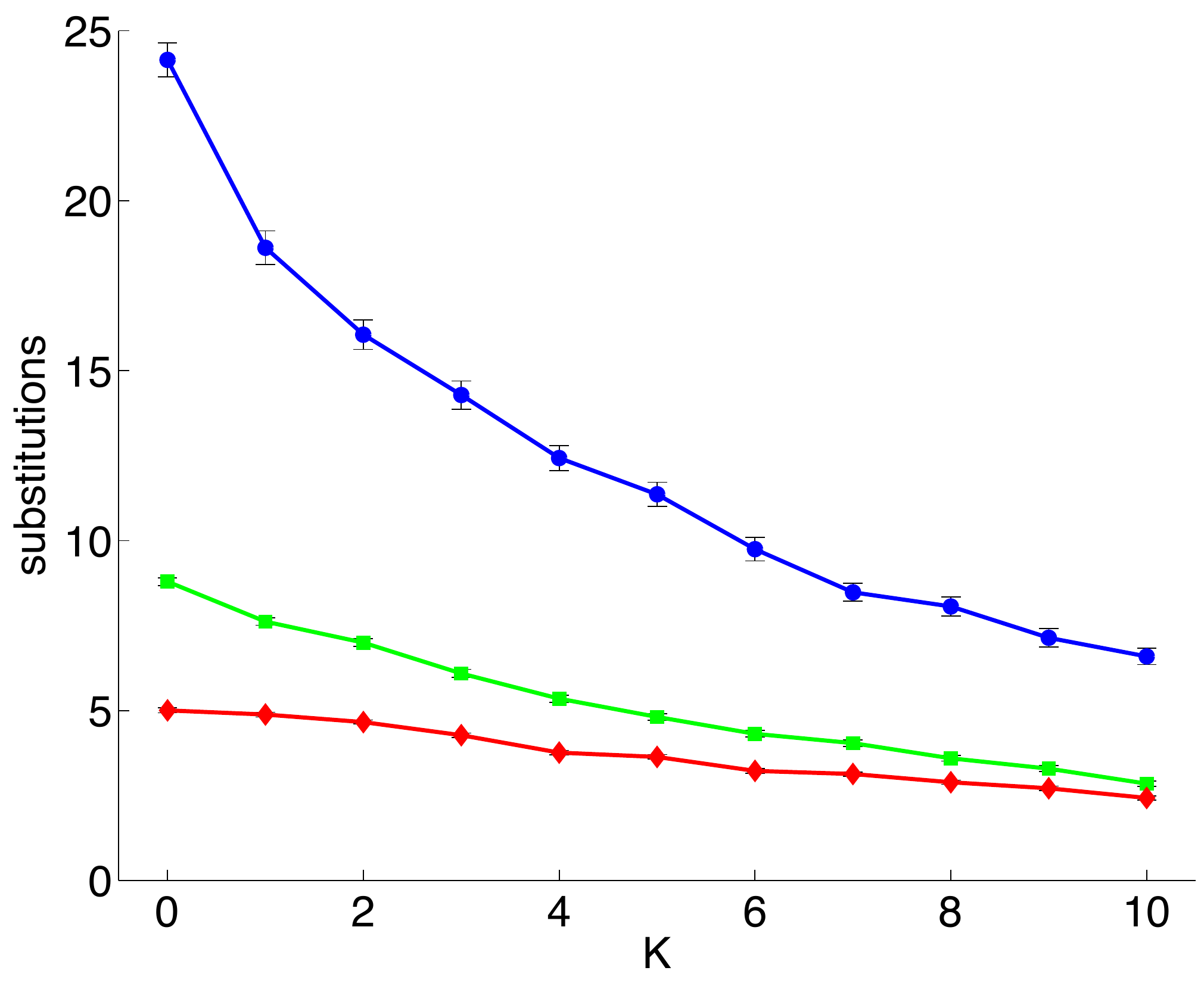}
\caption{The number of substitutions declines as the ruggedness of the landscapes increases. Red diamonds: $\mu=10^{-4}$ , green squares: $\mu=10^{-3}$, blue circles: $\mu=10^{-2}$. Data are averages over 200 simulation runs. Error bars are s.e.m. From~\cite{ostman2012}.}
\label{fig_2}
\end{center}
\end{figure}

Because the objective is to study the adaptive phase of evolutionary dynamics (as opposed to mutation-selection balance~\cite{Desai2007}), the simulations are started with a population of low fitness, allowing the population to increase in fitness. Every computational update 10\% of 5,000 asexual organisms are removed at random, and the remaining organisms replace them by reproduction. The organisms that get to reproduce are chosen with a probability proportional to fitness. That is, if the fitness of an organism is twice that of another, it has twice the chance to reproduce. Since this process is stochastic, organisms of lower fitness are not doomed to extinction, but with luck can become the ancestors of later generations. Reproduction is simulated by making a copy of the chosen organism and allowing each of the $N$ loci to mutate from 0 to 1 or from 1 to 0 at a rate $\mu$. Three different mutation rates are used in order to study the effect of the mutation-supply rate on the ability of the populations to cross valleys in the fitness landscape. Every organism has a genotype that consists of a binary string of length $N=20$ loci, and the effect of varying degrees of ruggedness on adaptation is investigated by running simulations with different values of $K$, which modulates the amount of epistatic interactions. Each simulation is run for 2,000 updates, which is enough in most instances to attain the highest fitness possible given the fitness landscape, population size, and mutation rate. Because there are no features of the dynamics that allow more than transient coexistence of different genotypes, the most recent common ancestor is never far in the past. Consequently, all organisms surviving at the end of a simulation run shares most of their history, and we can therefore reconstruct the line of descent (LOD) that is common to all surviving organisms and still cover most of the 2,000 computational updates. Epistasis is measured between pairs of consecutive mutations, $i$ and $j$, on this shared LOD in the following way:

\be
\epsilon_{ij} = \log \left( \frac{W_0W_{AB}}{W_AW_B} \right) ,
\ee
 
where $W_0$ denotes fitness before either mutation, $W_A$ is fitness with the first mutation, $W_B$ is fitness with the second mutation, and $W_{AB}$ is fitness with both mutations (Fig.~\ref{fig_1}). The genotype with only mutation B does not occur on the LOD, so $W_B$ has to be reconstructed afterwards for epistasis to be calculated. Epistasis on the LOD is calculated as the average of each pair of mutations. Epistasis can readily be calculated both between non-consecutive mutations as well as between more than two mutations, but we are here interested in how epistatic interactions enable populations to cross fitness valleys, and the most recent mutations have the largest effect on the ability to do this.

\begin{figure}[htp]
\begin{center}
\includegraphics[width=4.3in,angle=0]{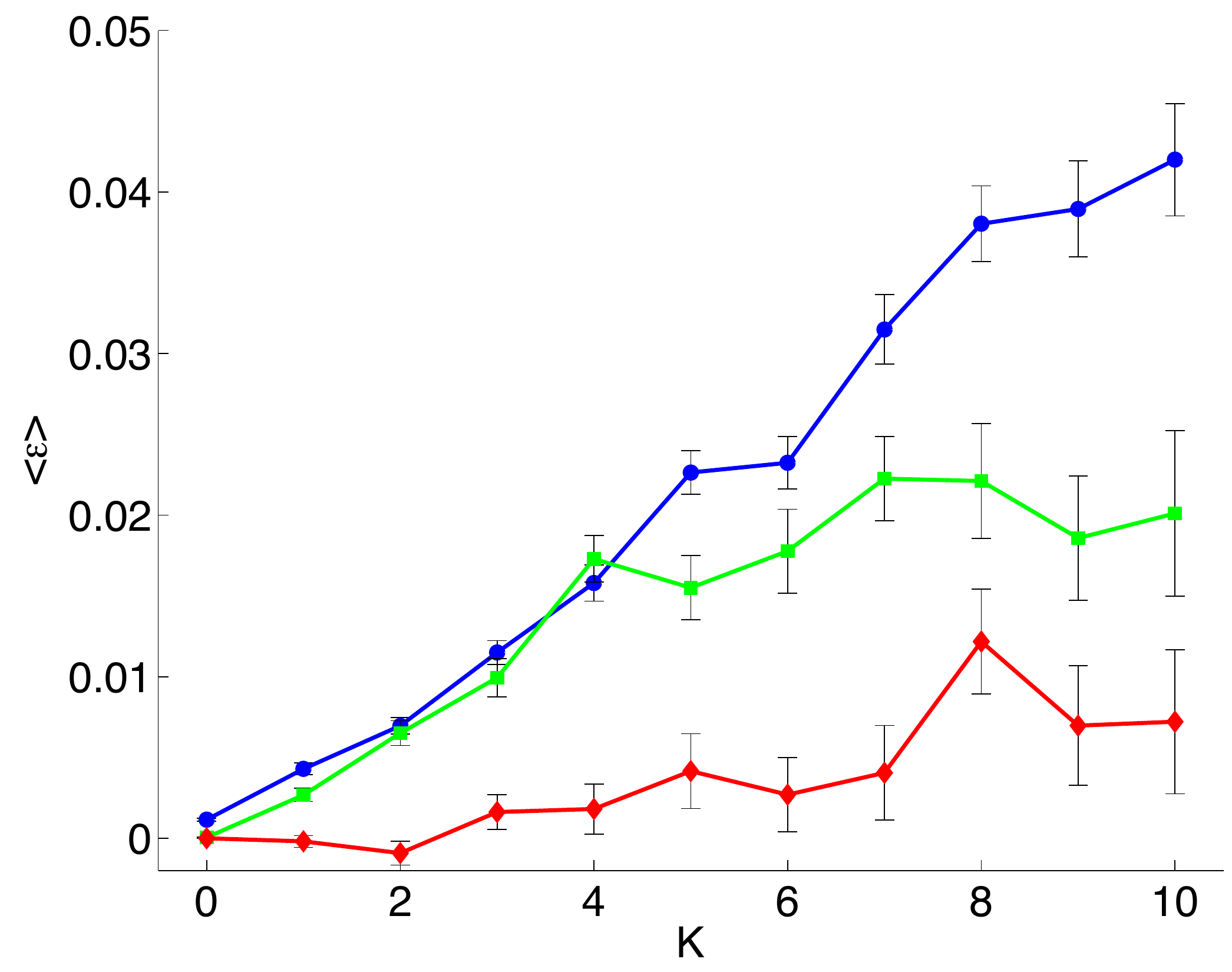}
\caption{Mean epistasis of all pairs of substitutions increases with landscape ruggedness. Colors and data as in Fig. 2. From~\cite{ostman2012}.}
\label{fig_3}
\end{center}
\end{figure}

\begin{figure}[htp]
\begin{center}
\includegraphics[width=4.3in,angle=0]{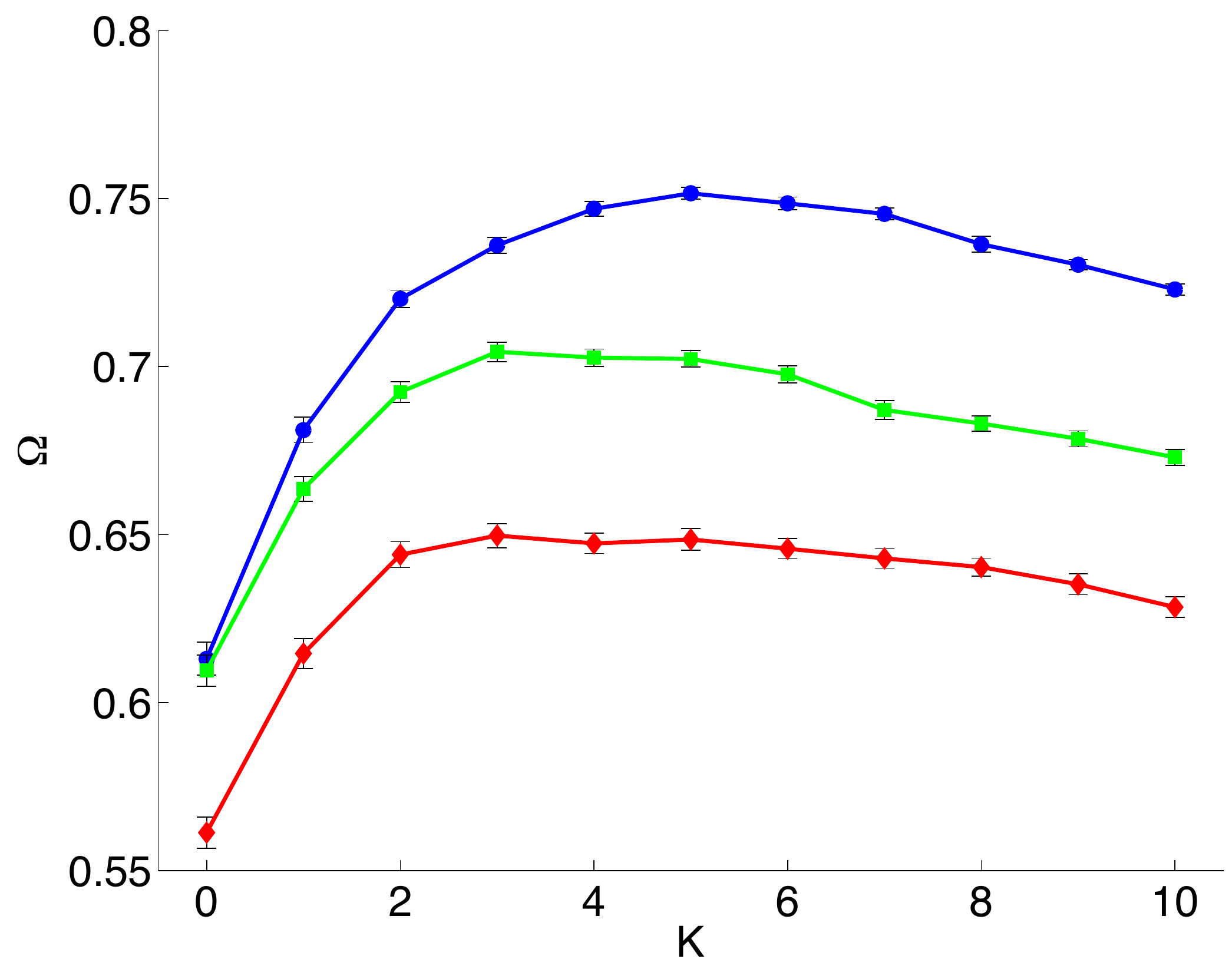}
\caption{Attained fitness increases with landscape ruggedness to a point after which a decline is observed. Colors as in Fig. 2. From~\cite{ostman2012}.}
\label{fig_4}
\end{center}
\end{figure}

\begin{figure}[htp]
\begin{center}
\includegraphics[width=4.3in,angle=0]{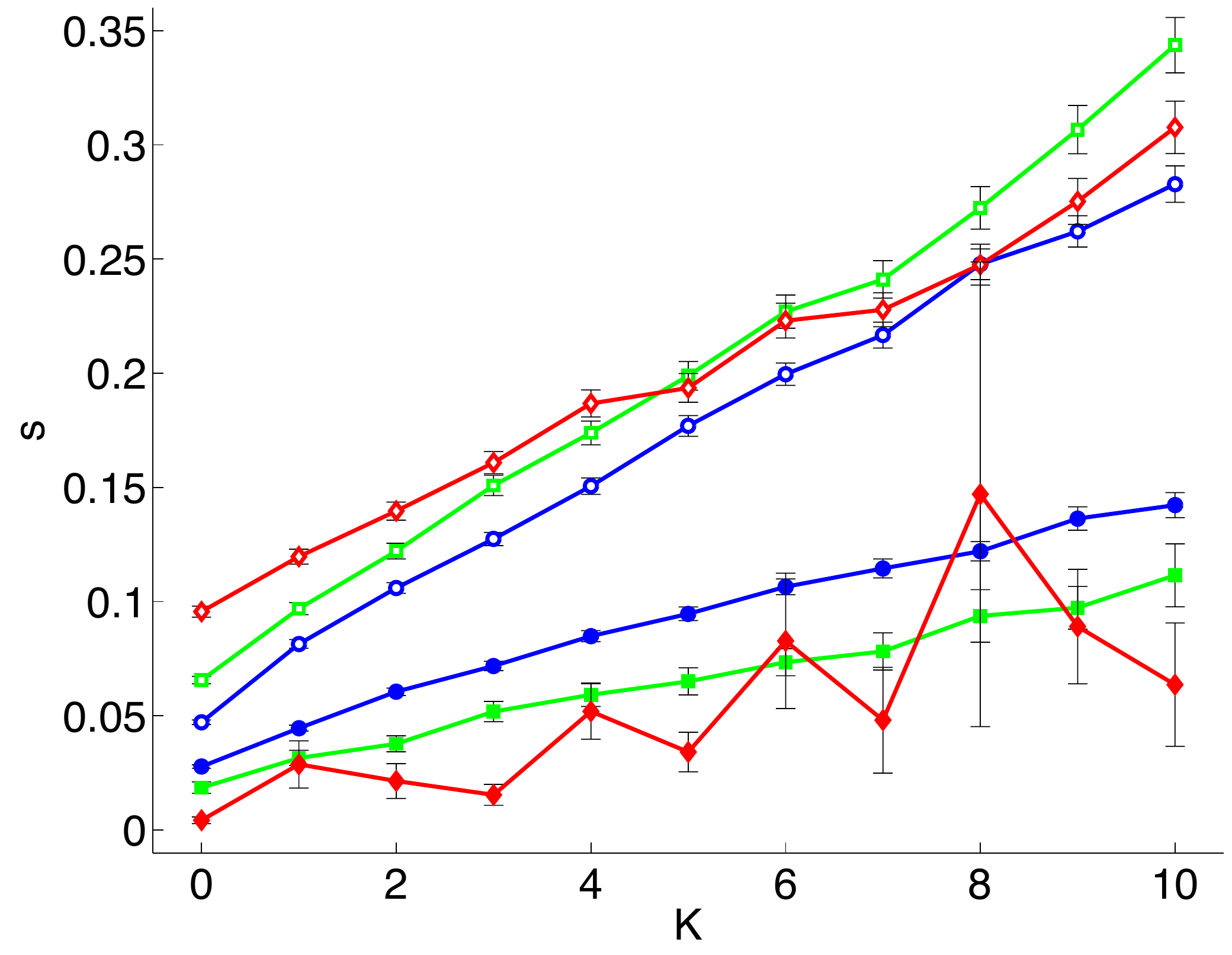}
\caption{Selection coefficients increase with landscape ruggedness. Colors as in Fig. 2. Open symbols are beneficial substitutions, and solid symbols are deleterious substitutions. From~\cite{ostman2012}.}
\label{fig_5}
\end{center}
\end{figure}

\begin{figure}[htp]
\begin{center}
\includegraphics[width=4.3in,angle=0]{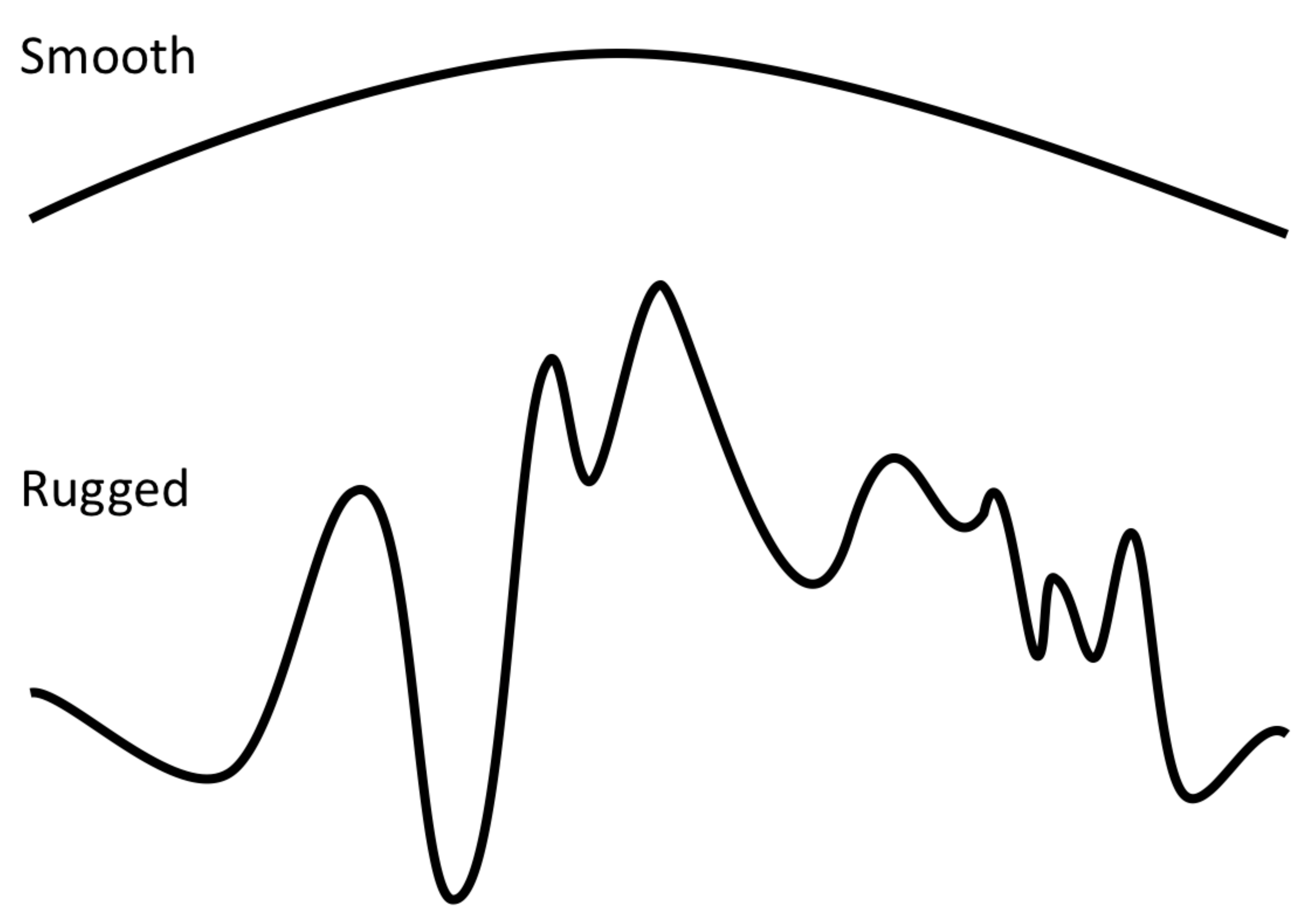}
\caption{Examples of smooth and rugged fitness landscapes in one dimension. The smooth landscape has a single peak whereas the rugged landscape has multiple local peaks. The ruggedness is caused by epistatic interactions between genes or mutations, and because epistasis and pleiotropy are coupled in the NK model, increased epistasis also leads to higher levels of pleiotropy. Ruggedness therefore implies not only more peaks but also a greater range in fitness, here represented by the scale of the vertical axes.}
\label{fig_6}
\end{center}
\end{figure}

\section{Results}
The NK model was used to investigate the extent to which adapting populations make use of epistatic interactions. Simulations were carried out for three different mutations rates of  $\mu=10^{-4}$,  $\mu=10^{-3}$, and $\mu=10^{-2}$. With a constant population size of 5,000, this gives mutation-supply rates of 0.5, 5, and 50, which spans values on either side of the limit demarcating the strong-selection weak-mutation regime (SSWM) from the regime where multiple mutations go to fixation at the same time. This regime has been investigated analytically under the assumption that the mutations do not interact~\cite{Unckless2009}, but here this assumption is removed allowing for epistasis to affect the adaptive process. Adaptation is observed and the fitness attained at the end of the 2,000 updates is recorded as $\Omega$, and used to compare the efficiency with which different populations adapt.

As $K$ is increased from 0 (smooth landscape with no epistasis) to 10 (highly rugged landscape), the number of substitutions, i.e., mutations on the LOD, drops significantly (Fig.~\ref{fig_2}). The fraction of substitutions that are beneficial stays approximately constant across this range of $K$, while the amount of epistasis on the LOD increases (Fig.~\ref{fig_3}). The attained fitness increases up to $K \approx 5$ (depending on the mutation rate), after which a decline is observed (Fig.~\ref{fig_4}). So despite there are fewer beneficial substitutions for higher $K$, the population is still able to attain a higher fitness. This seems contradictory, for surely more beneficial substitutions would result in higher fitness. The answer is that the structure of the fitness landscape is the main determinant of the extent to which an evolving population can adapt. In the NK model epistasis causes changes in fitness that deviate from the non-epistatic expectation. The effect of each mutation is modulated, and the selection coefficients increase with $K$ (Fig.~\ref{fig_5}). This increase in selection coefficients is an effect of both epistasis and pleiotropy. Epistasis causes the fitness landscape to be more rugged and therefore to contain more peaks, which will then have steeper slopes. Pleiotropy increases the effect a mutation can have on fitness by affecting more than one trait at a time. Both of these cause the distribution of selection coefficients to be broader, opening up more opportunities for mutations of large effect.

\section{Discussion}
\subsection{Epistasis Causes Landscape Ruggedness}

Fitness landscapes that contain only a single peak are often called smooth, whereas landscapes with multiple peaks are denoted as rugged (Fig.~\ref{fig_6}). Rugged landscapes can vary both in the number of peaks and in the range of fitness values between the least fit genotypes and the fitness of the global peak. This ruggedness of the fitness landscape is caused entirely by epistasis, with selection determining the fitness of individual genotypes. When there is no epistasis, the landscape is smooth. In this case, adaptation is straightforward, as an evolving population will eventually reach the top of the peak. When more than one peak is present, there is always epistasis present as well~\cite{Poelwijk2011}. Take for example a fitness landscape in one dimension where fitness is a function of body-size. If small and large bodies are favored by selection but intermediate sizes have lower fitness, then there are two peaks. Moving from one peak to the other necessarily results in epistasis between at least one pair of mutations: at the bottom of the valley, one mutation causes an increase in fitness only on the background of another, without which there would be a decrease in fitness.

While fitness landscape ruggedness is caused by epistasis by increasing the number of peaks, it is pleiotropy that is responsible for the increased range in fitness values. If we imagine the fitness landscape as a one-dimensional string, then the smooth landscape is simply a string with a single peak (Fig.~\ref{fig_6}). It has a range from one of the ends of the string to the height of the global peak. If we increase the number of epistatic interactions between genes and mutations, then the string becomes wrinkled, resulting in multiple local peaks. However, if we were to do this without affecting pleiotropy, then the fitness range would be unaffected. On the other hand, if we only increase the average level of pleiotropy while keeping the number of epistatic interactions constant, then the fitness range would expand without affecting ruggedness. This happens because pleiotropy results in mutations affecting more than one trait at a time, thereby broadening the distribution of the fitness effect of mutations, whether deleterious or beneficial. We can express this as epistasis modulates the frequency of the landscape, while pleiotropy modulates the amplitude. In the NK model epistasis and pleiotropy are coupled, and K directly affects both at the same time. In natural systems this is generally not the case. Rather, genes group into modules that affect individual phenotypic traits, and through their joint action on this trait the genes (an mutations affecting them) interact epistatically. In contrast, while genes generally interact with many other genes, the level is pleiotropy is neither tied to epistasis nor is it as prevalent. The emerging picture from natural systems (e.g., yeast) is that there are relatively few pleiotropic links between modules, resulting in well-defined modules of genes affecting single traits~\cite{Costanzo2010}. Consequently, the effects of epistasis and pleiotropy are not coupled in natural systems, and decoupling them in the NK model could therefore increase the realism of the model. If the level of pleiotropy is lower in natural systems, then the range in fitness will also be affected, and the linear relationship found between selection coefficients and K will probably be less steep.

\subsection{Future directions}
It is generally assumed that the fitness effects of mutations in pleiotropic genes are uncorrelated (e.g.,~\cite{Orr1998}). Because most mutations are either neutral or deleterious, a mutation that is beneficial in one trait is considered most likely to be neutral or deleterious in the other traits that the pleiotropic gene affects. However, considering that the gene encodes a protein that is likely to have the same function in several or all the traits, it seems more likely that if the mutation is beneficial for one trait, then it would also likely be beneficial in the linked traits. For example, if the protein has the same function in the linked traits, then mutations that improve thermal stability or affinity of the active site are likely to be beneficial in all the traits. If the function of the protein is different in the linked traits, then there may be no correlation between the fitness effects, but for proteins whose biochemical function is similar in the linked traits, a correlation in fitness is likely. Such pleiotropically correlated fitness effects could have consequences for adaptation, causing beneficial mutations in pleiotropic genes to have larger effect, thereby increasing both the speed and probability of fixation.

In the weak mutation regime, early adaptation in NK is dominated by non-interacting beneficial mutations with some negative epistasis. Later in the adaptive process and nearer the peak epistasis shifts to become predominantly positive and to include some sign epistasis~\cite{Draghi2012}. The diminishing returns observed among beneficial mutations in adapting microbial populations~\cite{Chou2011,Khan2011} thus appears to be due to regression to the mean. Experimental populations are generally not seen crossing valleys in the fitness landscape, so the positive and reciprocal sign epistasis that would then be observed have rarely been reported in the literature (but see~\cite{Dawid2010,Kvitek2011}). Valleys can be crossed when the mutation-supply rate is large enough, and the waiting time to new mutations is short and deleterious mutations can be tolerated. Reciprocal sign epistasis is a necessary condition for multi-peaked fitness landscapes~\cite{Poelwijk2011}, so empirical observations would therefore shed light on this important aspect of fitness landscape structure.

In the NK model, global peak height increases with landscape ruggedness~\cite{ostman2012}. However, this may not be an artifact of the NK model alone. Rather, because the synergy between genes in modules is dependent on the number of genes, modules of genes encoding traits will have a larger effect on fitness the more genes are available. In other words, larger modules confer higher fitness. A comparison between different species that share traits (e.g. vision) should reveal a correlation between the fitness conferred by the module and the number of genes in the module. This could be measured by counting the number of genes that contribute to, say, vision among similar organisms and scoring the trait on a scale of how well it functions. Most likely the increase in fitness as a function of the number of genes is less than linear, because the function cannot be improved indefinitely.

\bibliography{EBM16}

\begin{thebibliography}{10}
\providecommand{\url}[1]{{#1}}
\providecommand{\urlprefix}{URL }
\expandafter\ifx\csname urlstyle\endcsname\relax
  \providecommand{\doi}[1]{DOI \discretionary{}{}{}#1}\else
  \providecommand{\doi}{DOI \discretionary{}{}{}\begingroup
  \urlstyle{rm}\Url}\fi

\bibitem{Gavrilets2004}
S.~Gavrilets, \emph{{Fitness Landscapes and the Origin of Species}} (Princeton
  University Press, 2004)

\bibitem{Whitlock1995}
M.C. Whitlock, P.C. Phillips, F.B.G. Moore, S.J. Tonsor, Annual Review of
  Ecology and Systematics \textbf{26}, 601 (1995)

\bibitem{Mustonen2009}
V.~Mustonen, M.~L{\"a}ssig, Trends in Genetics \textbf{25}(3), 111 (2009)

\bibitem{Whitlock1997}
M.C. Whitlock, Evolution \textbf{51}(4), 1044 (1997)

\bibitem{Wright1932}
S.~Wright, Proceedings of the Sixth International Congress of Genetics
  \textbf{1}, 356 (1932)

\bibitem{Gavrilets1997}
S.~Gavrilets, Trends in Ecology {\&} Evolution \textbf{12}(8), 307 (1997)

\bibitem{Kauffman1993}
S.~Kauffman, \emph{{The Origins of Order}} (Oxford University Press, New York,
  1993)

\bibitem{Kauffman1987}
S.~Kauffman, S.~Levin, Journal of Theoretical Biology \textbf{128}(1), 11
  (1987)

\bibitem{ostman2012}
B.~{\O}stman, A.~Hintze, C.~Adami, Proceedings of the Royal Society B:
  Biological Sciences \textbf{279}, 247 (2012)

\bibitem{Desai2007}
M.M. Desai, D.S. Fisher, Genetics \textbf{176}(3), 1759 (2007)

\bibitem{Unckless2009}
R.L. Unckless, H.A. Orr, Genetics \textbf{183}, 1079 (2009)

\bibitem{Poelwijk2011}
F.J. Poelwijk, S.~T{\u a}nase-Nicola, D.J. Kiviet, S.J. Tans, Journal of
  Theoretical Biology \textbf{272}(1), 141 (2011)

\bibitem{Costanzo2010}
M.~Costanzo, A.~Baryshnikova, J.~Bellay, Y.~Kim, E.D. Spear, C.S. Sevier,
  H.~Ding, J.L.Y. Koh, K.~Toufighi, S.~Mostafavi, J.~Prinz, R.P. St~Onge,
  B.~Vandersluis, T.~Makhnevych, F.J. Vizeacoumar, S.~Alizadeh, S.~Bahr, R.L.
  Brost, Y.~Chen, M.~Cokol, R.~Deshpande, Z.~Li, Z.Y. Lin, W.~Liang,
  M.~Marback, J.~Paw, B.J. San~Luis, E.~Shuteriqi, A.H.Y. Tong, N.~van Dyk,
  I.M. Wallace, J.A. Whitney, M.T. WEirauch, G.~Zhong, H.~Zhu, W.A. Houry,
  M.~Brudno, S.~Ragibizadeh, B.~Papp, C.~Pal, F.P. Roth, G.~Giaever, C.~Nislow,
  O.G. Troyanskaya, H.~Bussey, G.D. Bader, A.C. Gingras, Q.D. Morris, P.M. Kim,
  C.A. Kaiser, C.L. Myers, B.J. Andrews, C.~Boone, Science \textbf{327}, 425
  (2010)

\bibitem{Orr1998}
H.A. Orr, Evolution \textbf{52(4)}, 935 (1998)

\bibitem{Draghi2012}
J.A. Draghi, J.B. Plotkin, arXiv:1212.4114v1 [q-bio.PE]  (2012)

\bibitem{Chou2011}
H.~Chou, H.~Chiu, N.~Delaney, D.~Segr{\`e}, C.J. Marx, Science \textbf{332},
  1190 (2011)

\bibitem{Khan2011}
A.I. Khan, D.M. Dinh, D.~Schneider, R.E. Lenski, T.F. Cooper, Science
  \textbf{332}, 1193 (2011)

\bibitem{Dawid2010}
A.~Dawid, D.J. Kiviet, M.~Kogenaru, M.~de~Vos, S.J. Tans, CHAOS \textbf{20}
  (2010)

\bibitem{Kvitek2011}
D.J. Kvitek, G.~Sherlock, PLoS Genet \textbf{7}(4), e1002056 (2011)

\end{thebibliography}
\bibliographystyle{spphys}
\end{document}